\documentstyle[11pt,aaspp4]{article}

\newcommand\etal{et~al.}
\newcommand\cm{\hbox{cm}}
\newcommand\s{\hbox{s}}
\newcommand\gm{\hbox{gm}}
\newcommand\ergs{\hbox{ergs}}
\newcommand\keV{\hbox{keV}}
\newcommand\MeV{\hbox{MeV}}

\newcommand\Hz{\hbox{Hz}}
\newcommand\pc{\hbox{pc}}
\newcommand\pf{Phys. Fluids}

\lefthead{Brainerd}
\righthead{A Theory of Gamma-Ray Burst Emission}

\begin{document}

\title{A Plasma Instability Theory of Gamma-Ray Burst Emission}

\author{J. J. Brainerd\altaffilmark{1}}

\affil{University of Alabama in Huntsville}

\altaffiltext{1}{Space Sciences Lab, ES-84, NASA/Marshall Space Flight Center,
Huntsville, AL \ 35812\\
Jim.Brainerd@msfc.nasa.gov\\}

\begin{abstract}
A plasma instability theory is presented for the prompt radiation from
gamma-ray bursts.  In the theory, a highly relativistic shell interacts
with the interstellar medium through the filamentation and the
two-stream instabilities to convert bulk kinetic energy into electron
thermal energy and magnetic field energy.  The processes are not
efficient enough to satisfy the Rankine-Hugoniot conditions, so a
shock cannot form through this mechanism.  Instead, the interstellar
medium passes through the shell, with the electrons radiating during
this passage.  Gamma-rays are produced by synchrotron self-Compton
emission.  Prompt optical emission is also produced through this
mechanism, while prompt radio emission is produced through synchrotron
emission.  The model timescales are consistent with the shortest burst
timescales.  To emit gamma-rays, the shell's bulk Lorentz factor must be
$\gtrsim 10^3$.  For the radiative processes to be
efficient, the interstellar medium density must satisfy a lower limit
that is a function of the bulk Lorentz factor.  Because the limits
operate as selection effects, bursts that violate them constitute new
classes.  In particular, a class of optical and ultraviolet bursts with
no gamma-ray emission should exist.  The lower
limit on the density of the interstellar medium is consistent
with the requirements of the Compton attenuation theory, providing
an explanation for why all burst spectra appear to be attenuated. 
Several tests of the theory are discussed, as are the next theoretical
investigations that should be conducted.
\end{abstract}

\keywords{Gamma rays: bursts}


\section{Overview}

Current observations of gamma-ray bursts place a number of strong
constraints on gamma-ray burst physics.  The measured red shift of
$z = 0.835$, $0.966$, and $1.61$ for lines in the optical emission
of gamma-ray bursts GRB~970508
(Metzger \etal\ \markcite{Metzger1}1997a,\markcite{Metzger1}b), 
GRB~980703
(Djorgovski \etal\ \markcite{Djorgovski1}1998a,\markcite{Djorgovski2}b), 
and GRB~990123 (\markcite{Kelson}Kelson \etal\ 1999), 
respectively, and the indirect redshifts of $z = 3.5$, $5$, and $1.096$
for GRB~971214 (\markcite{Kulkarni}Kulkarni, S. R., \etal\ 1998), 
GRB~980329 (\markcite{Fruchter}Fruchter 1999), 
and GRB~980613 (\markcite{Djorgovski3}Djorgovski \etal\ 1999), 
respectively, show that gamma-ray bursts are at extraordinarily-high
redshifts.  The power-law gamma-ray spectrum
above $511 \, \keV$ and the rapid rise times of the gamma-ray light-curves
force one to consider theories with highly-relativistic bulk motion in order
to avoid thermalization of the gamma-rays through photon-photon pair creation
(\markcite{Schmidt}Schmidt 1978;
\markcite{Baring}Baring, \& Harding 1996). 
The gamma-ray spectrum can be characterized by the E-peak energy ($E_p$),
which is the photon energy of the maximum of the $\nu F_{\nu}$ spectrum.
Because the distribution of $E_p$ values is narrow, with an average value
of $E_p \approx 250 \, \keV$
(\markcite{Mallozzi}Mallozzi \etal\ 1996;
\markcite{Brainerd4}Brainerd \etal\ 1999), 
one suspects that the characteristics of the spectrum are independent
of the bulk Lorentz factor, which should vary greatly from burst to burst.
In some bursts, the shape of the gamma-ray spectrum is inconsistent
with optically thin synchrotron emission
(Preece \etal\ \markcite{Preece2}1998a,\markcite{Preece3}b). 
The shape of the burst spectrum is consistent with being
Compton attenuated by a high density ($\approx 10^5 \, \cm^{-3}$)
interstellar medium
(\markcite{Brainerd1}Brainerd 1994;
Brainerd \etal\ \markcite{Brainerd2}1996,\markcite{Brainerd3}1998);
observations show the x-ray excess predicted by
this theory (\markcite{Preece4}Preece \etal\ 1996), 
and the redshifts of $\approx 1$ to $\approx 10$
derived by fitting the model to burst spectra are consistent with
the values measured at optical wavelength
(\markcite{Preece1}Preece \& Brainerd 1999). 
The success of this theory, its presence in every gamma-ray burst,
implies that a high density interstellar medium is necessary
for a burst to occur.

The observations suggest that the sources of gamma-ray bursts are
compact---perhaps a several mass black hole, perhaps a supermassive
black hole---and that these sources eject mass at relativistic velocities
in one short event.  The observed behavior of the gamma-ray burst arises
from processes that convert the kinetic energy of the ejected material
into electromagnetic radiation.

The general view that has developed concerning the gamma-ray emission
of gamma-ray bursts is that it is radiated behind a shock that has
developed within a shell moving with a bulk Lorentz factor of
$\Gamma \approx 10^3$ (\markcite{Meszaros}M\'esz\'aros 1998). 
The radiative mechanism universally cited is synchrotron emission
from electrons accelerated to an energy close to the equipartition
energy (\markcite{Tavani}Tavani 1996).  The shock itself is collisionless,
arising either when the shell runs into the interstellar medium, or
when the fastest portions of the shell overtake slower portions
(Rees \& Mesaros \markcite{Rees1}1992, \markcite{Rees2}1994;
\markcite{Piran}Piran \& Sari 1998;
\markcite{Mochkovitch}Mochkovitch \& Daigne 1998).
This model has several shortcomings.  First, the gamma-ray burst
spectrum is often harder at x-ray energies than is allowed by an
optically-thin synchrotron emission model.
Second, it provides no explanation for the x-ray excess, for
the narrow $E_p$ distribution, or for the apparent absence of
bursts with $\Gamma < 10^3$, as inferred from the absence of
photon-photon pair creation and thermalization.
From the theoretical side, many assumptions are made without
theoretical development; among these are that collective processes exist
that mediate the shock, that a strong magnetic field is generated
in the shock, and that the energy is efficiently transferred from
the bulk-motion of the ions to the thermal energy of the electrons.

It is with these difficulties in mind that a new theory for the
generation of the prompt radiation in a gamma-ray burst
is proposed.  The theory is a plasma instability theory.  In this
theory, the shell of relativistic material that is ejected from the source
interacts with the interstellar medium through the plasma filamentation
instability, which, in the relativistic regime, has a higher growth rate
than the two-stream instability.  Because the mass ejection event
must be short, of order the
burst durations, and because the mass travels $2 c \Gamma^2$ times the
burst duration, the interaction is of a thin shell with the interstellar
medium.  As the shell passes through the interstellar medium, the
interstellar medium collapses into filaments that contain strong magnetic
fields and high electron thermal energies.  It is the interstellar medium that
filaments rather than the shell, because the interstellar medium has the
smaller density as measured in each comoving reference frame.  The
ions in the filaments remain essentially at rest with respect
to the observer, while the electrons move towards the observer with
a high bulk Lorentz factor.  The magnetic fields generated through
filamentation are strong enough to produce gamma-rays through synchrotron
self-Compton emission.  One finds that the theory places lower limits
on both $\Gamma$ and the interstellar medium density through selection
effects, and that these lower limits lead to the conditions required
by the Compton attenuation theory.  The theory implies that there
exists a class of burst that produces intense and prompt optical and
ultraviolet emission, but no x-rays and gamma-rays.

In this article, I give an analytic development of the theory outlined
above.  This development explores the relevant plasma and radiative
processes, deriving the selection effects inherent in the theory,
and ascertaining the aspects of the theory that provide means of
observationally testing the theory.  In \S 2, the characteristic
ratio of shell density to interstellar medium density and the
characteristic thickness of the shell are discussed.  In \S 3, the
growth of the two-stream and filamentation plasma instabilities
is examined.  The saturation of the filaments and the inability
of the filamentation instability to mediate a shock are examined in \S 4.
The electron thermalization and isotropization and the rest frame
defined by the electron component are discussed in \S 5.
The radiative processes of synchrotron emission and
synchrotron self-Compton emission are examined in \S 6, where
the characteristic frequencies and emission rates of each
are derived.  In \S 7, the radiative timescales for synchrotron
and synchrotron self-Compton cooling are derived.  The theory
has several natural selection effects that constrain several of
the free parameters in the theory.  These are discussed in \S 8.
The basic theory and the results of this study are summarized in \S9.
This section also contains some suggestions for observational
tests of the validity of the theory.


\section{Characteristic Model Parameters}

The model is of a fully-ionized neutral shell of electrons
and protons passing through a fully-ionized interstellar medium
of electrons and protons.  The characteristic physical parameters
that describe this theory are $n_{ism}$,
the number density of the electrons or of the protons in the
interstellar material, $n_{shell}^{\prime}$, the number density
of the electrons or protons in the relativistic shell, and $\Gamma$,
the Lorentz factor of the shell relative to the interstellar medium.
The quantities with primes are measured in the shell rest frame, while
those without primes are measured in the interstellar medium rest
frame.  An important parameter in the discussion that follows is
the parameter $\eta$, defined as
\begin{equation}
   \eta = { n_{ism} \over n_{shell}^{\prime} }
   \, .
\end{equation}

There is no simple connection between $\eta$, $\Gamma$, and the
parameters that describe the energetics of the system.  In
particular,  to relate $\eta$ to ${\cal M}$, the mass per unit ster
radian of the relativistic shell, and $R$, the distance traveled
before deceleration, requires a good understanding
of the radiative transfer and plasma physics of the problem.
The value of $\eta$ is therefore treated as a free parameter,
with the limits on the acceptable values of $\eta$ set by
a combination of theoretical and observational constraints.
The former is set in this section, while the latter is set in \S 8. 

A specific value of $\eta$ that has a physical significance is
the value of $\eta$ expected from the continuity equation for
a relativistic shock.  For $\Gamma \gg 1$, one has
$n_{shell}^{\prime} = n_{ism} \Gamma$, and
\begin{equation}
   \eta_{shock} = { 1 \over \Gamma}
   \, .
\end{equation}
When $\eta < \eta_{shock}$, the interstellar medium
density in the shell rest frame is less than the shell density,
while it is greater than the shell density when $\eta > \eta_{shock}$.

A lower limit $\eta_{min}$ on $\eta$ can be derived through
a theoretical argument concerning pressure equilibrium with
in the shell.  If one assumes that the interaction with
the interstellar medium exerts a force only at the front
of the shell, then one can relate $\eta$ to the temperature
of the shell and $R$, the distance the shell has traveled
from the source when $\Gamma$ is a factor of 2 below its initial
value.  This defines a maximum value for the shell density,
and therefore a minimum density for $\eta$, because when
the deceleration force is spread throughout the shell,
a lower pressure is required to maintain static equilibrium
within the shell.  Defining the dimensionless 4-velocity
in the radial direction as $u_r$, the deceleration of the shell
is given by
\begin{equation}
   { {\cal M} c \over R^2 } { d u_r \over d \tau }
   = \Gamma p^{\prime} = \Gamma n_{shell,max}^{\prime} T
   \, .
\end{equation}
In this equation, the pressure $p^{\prime}$ is the pressure exerted
on the shell by the interaction with the interstellar medium as measured
in the shell rest frame.  The factor of $\Gamma$ converts this proper
measure of force into the radial component of the 4-acceleration.
The parameter $T$ is the sum of the electron and proton temperatures
in the shell in units of energy.  Changing
the derivative on the left-hand side of equation (3) into a derivative
in $r$ and setting $dr \approx R$ and $d u \approx \Gamma$ gives
a maximum density of
\begin{equation}
   n_{shell,max}^{\prime} \approx { {\cal M} c^2 \Gamma  \over 2 R^3 T }
   \, .
\end{equation}
The value of $R$ in this equation is bounded by a
minimum distance $R_0$
that is set by the case in which the interstellar medium is swept
up by the shell.  The amount of interstellar
medium per unit ster radian that must be
swept up to change $\Gamma$ by a factor of 2 is
$m \approx {\cal M}/\Gamma$, so
\begin{equation}
   R_0 = \left( {3 {\cal M} \over m_p n_{ism} \Gamma } \right)^{1/3}
   = 3.94 \times 10^{-3} \, \pc \; {\cal M}_{27}^{1/3} n_{ism}^{-1/3}
   \Gamma_3^{-1/3}
   \, ,
\end{equation}
where ${\cal M}_{27}$ is the shell mass per unit ster radians in
units of $10^{27} \gm$, $n_{ism}$ is given in units of $\cm^{-3}$,
and $\Gamma_3$ is the Lorentz factor in units of $10^{3}$.
With these parameters, a shell subtending $4 \pi$ ster radians
and having ${\cal M}_{27} = n_{ism} = \Gamma_3 = 1$ will
carry $1.129 \times 10^{52} \ergs$ of energy, which is the characteristic
value inferred from the observations at gamma-ray energies.
Using equation (5) to express $R$ in units of $R_0$ in equation (4) gives
\begin{equation}
   n_{shell,max}^{\prime} = { m_p c^2 n_{ism} \Gamma^2  \over 6 T }
   \, \left( {R_0 \over R} \right)^3 \, .
\end{equation}
The ratio of $n_{ism}$ to $n_{shell}^{\prime}$ is then
\begin{equation}
   \eta_{min} = {n_{ism} \over n_{shell,max}^{\prime} }
   = { 6 T \over m_p c^2 \Gamma^2 }
   \, \left( {R \over R_0} \right)^3 \, .
\end{equation}
The value of $\eta_{min}$ is strongly dependent on the distance
traveled.  For deceleration over $R = R_0$ with $T = m_p c^2 \Gamma$,
which is the temperature found when the interstellar
medium is swept up adiabatically,
one finds $\eta_{min} \approx \Gamma^{-1} = \eta_{shock}$,
as expected for a shock.  For lower values of $T$, one has
$\eta_{min} < \eta_{shock}$ unless $R > R_0$ by a sufficiently large
value.  For $T \approx m_e c^2 \Gamma$, which is the case for
the theory discussed below, then one has
$\eta < \eta_{shock}$ for $R < 6.7 \, R_0$.
Lower limits on $\eta$ for $T = m_e c^2 \Gamma$ and several
different values of $\Gamma$ are given as functions of $R/R_0$
in Figures 1 and~2.

\begin{figure}
\figurenum{1: Limits on $\eta$}
\epsscale{0.7}
\plotone{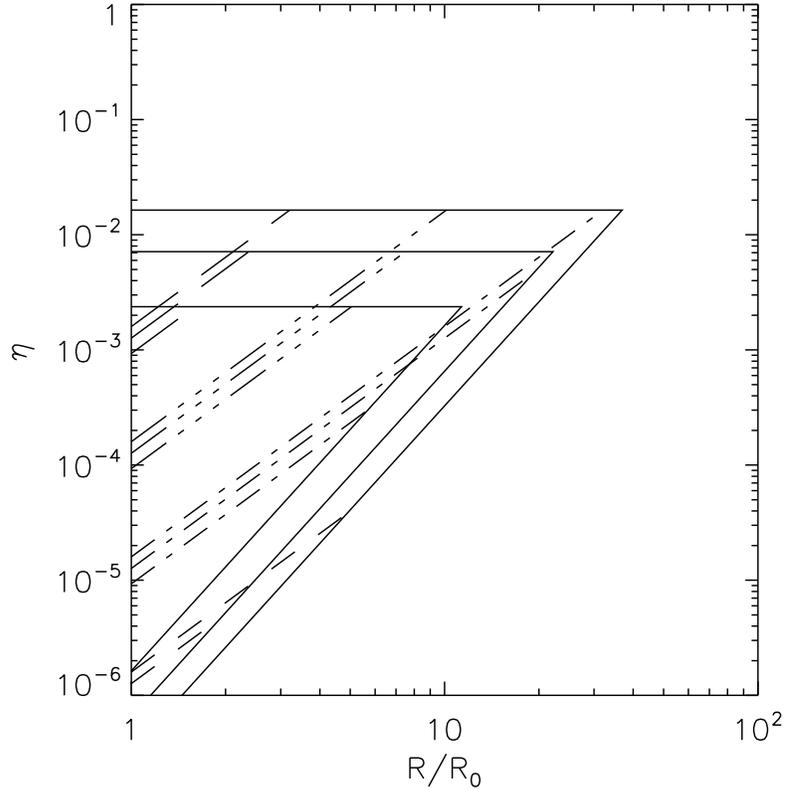}
\caption{
The variable $\eta$ is defined by equation (1) as the
ratio of the interstellar medium proper density
to the shell proper density.  The shell deceleration distance
$R$ is plotted in terms of $R_0$, which is defined in eq\hbox{.}~(5).
Limits on the value of $\eta$ are
plotted as solid curves for the values $\Gamma = 2 \times 10^3$,
$5 \times 10^3$, and $10^4$, with the lower limits given by
equation (7), and the upper limits by equation (2).
The values of $\eta$ and $R$ that produce the shell thickness
timescales (eq. [12]) of $0.1 \, \s$, $1 \, \s$, $10 \, \s$,
and $100 \, \s$, are plotted from bottom
to top as a long-dashed, dot-dashed, three-dot-dashed, and long-dashed
lines, with a line for each value of $\Gamma$, which terminates on
the proper boundary for that $\Gamma$.
Other free parameters were set to $n_{ism} = 1 \, \cm^{-3}$,
${\cal M}_{27} = 1$.}
\end{figure}

\begin{figure}
\figurenum{2: Limits on $\eta$}
\epsscale{0.7}
\plotone{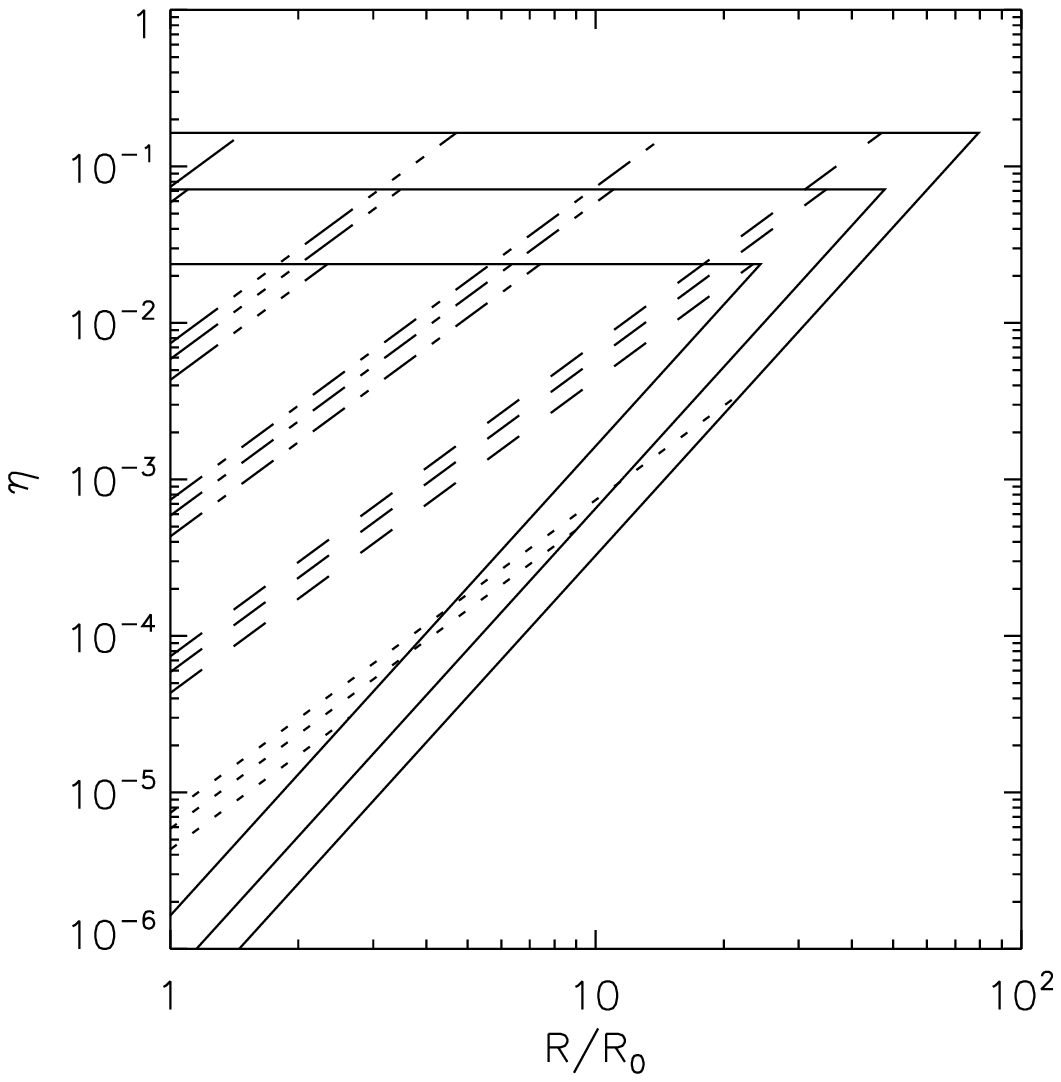}
\caption{The same as Fig\hbox{.}~1, except now
$n_{ism} = 10^{5} \, \cm^{-3}$.  The dotted-lines at the bottom of
the figure are for a shell thickness timescale of $0.01 \, \s$,
with the remaining curves the same as in Fig\hbox{.}~1.}
\end{figure}

The value of $\eta$ defines the light crossing time scale
across the shell width.  The thickness of the shell in
the shell rest frame is related to the mass of the shell by
\begin{equation}
   {{\cal M} \over R^2} = m_p n_{shell}^{\prime} l^{\prime}
   \, .
\end{equation}
Expressing $R$ in terms of $R_0$ through equation (5), one finds
\begin{eqnarray}
   l^{\prime} &=
   &\left( { {\cal M} \over m_p n_{ism} } \right)^{{1\over 3}} \,
   \left( {\Gamma \over 3 } \right)^{{2\over 3}} \,
   \left({R_0 \over R}\right)^2 \,
   \eta \, , \\
   &= &4.05 \times 10^{15} \, \cm \;
   {\cal M}_{27}^{{1\over 3}} \,
   n_{ism}^{-{1\over 3}} \,
   \Gamma_3^{{2\over 3}} \,
   \left( {R_0 \over R} \right)^2 \,
   \eta_{-3}
   \, .
\end{eqnarray}
where $\eta_{-3} = \eta/10^{-3}$.  The thickness of the shell in
the interstellar medium rest frame is $l = l^{\prime}/\Gamma$.
Defining the shell's characteristic timescale $t_{shell}$
as $l = c t_{shell}$, one has
\begin{eqnarray} t_{shell}
   &= &{\eta \over c }
   \left( { {\cal M} \over 9 m_p n_{ism} \Gamma} \right)^{{1\over 3}}
   \left({R_0 \over R}\right)^{2} \, , \\
   &= &1.35 \times 10^{2} \, \s \;
   {\cal M}_{27}^{{1\over 3}}
   n_{ism}^{-{1\over 3}} \Gamma_3^{-{1\over 3}}
   \left( {R_0 \over R} \right)^2 \eta_{-3}
   \, .
\end{eqnarray}
For $\eta$ of order $1/\Gamma$, one requires that
$R = 10 R_0$ at $n_{ism} = 1 \, \cm^{-3}$
for $t_{shell} = 1 \, \s$.  Equation (12) is plotted in Figures 1
and 2 as functions of $\eta$ with respect to $R/R_0$
for several values of $\Gamma$ and $t_{shell}$.
Because $t_{shell}$ must be less than or
equal to the burst timescales, which are often observed to be
$< 1 \, \s$, model values for $\eta$ must be $\eta \ll 1$.
It is shown in \S 8 that the theory provides this limit on $\eta$.


\section{Plasma Instabilities}

Within the reference frame of the shell, one initially has a plane-parallel
charge-neutral density profile through which a neutral and
uniform plasma streams.  The shell is assumed to have no initial magnetic
field.
The dissipation of the bulk kinetic energy of a relativistic shell to
the interstellar medium must be through a plasma instability, because
the densities of the shell and the interstellar medium ensure that the
collision mean free path is much longer than the thickness of the shell.
For a relativistic plasma, the two relevant instabilities are the two-stream
instability and the electromagnetic filamentation
instability (\markcite{Davidson1}Davidson 1990). 
For
$v \ll c$, the former has a larger growth rate that the latter by a factor
of $c/v$, but for $\Gamma \gg 1$, the growth rate of the latter becomes
much larger than the growth rate of the former.

The two-stream instability has a maximum growth rate in the relativistic
regime of
\begin{equation}
   \gamma_{2s}^{\prime} = {1 \over 2} \omega_{p, e, ism}^{\prime}
   \Gamma^{-{3 \over 2}}
   \approx {1 \over 2} \, \sqrt{ { 4 \pi e^2 \over m_e } }
   n_{ism}^{{1 \over 2}} \Gamma^{-1}
   \, ,
\end{equation}
where $\omega_{p, e, ism}^{\prime}$ is the electron plasma frequency
of the interstellar medium in the shell rest frame.
Taking $\Gamma = 10^3$ and $n_{ism} = 1 \, \cm^{-3}$,
one finds a growth rate for the electron two-stream instability of
$\gamma_{2s, e}^{\prime} = 30 \, \s^{-1}$ in the shell frame.  This growth
rate corresponds to distances of travel of $10^9 \, \cm$ for the interstellar
medium through the shell.

The filamentation instability acts on counterstreaming plasmas by creating
a magnetic pinch.  Of the two plasmas, the interstellar medium and the
shell, that with the smaller number density in it's own rest frame
is the plasma that filaments.
For gamma-ray bursts, this is the interstellar medium when $\eta \ll 1$.

The growth rates of the filamentation instability is easily derived from
the dielectric tensor for cold electron and ion streams in the relativistic
regime and in the absence of a magnetic field.  The equation must
satisfy (\markcite{Davidson1}Davidson 1990) 
\begin{equation}
   1 + \sum_{j} \left[
   { \omega^{2}_{pj} \over \Gamma_j^3 c^2 k_{\perp}^2 }
   + { \beta_j \omega^2_{pj} \over \Gamma_j \omega^2 } \right] = 0 \, ,
\end{equation}
where $\omega_{pj}$ is the plasma frequency of component $j$ for
the density measured in the observer's rest frame, $\Gamma_j$
is this component's Lorentz factor, and $k_{\perp}$
is the wave number perpendicular to the
velocity vector of the streams.

If one is considering only a charge neutral stream of electrons and
ions moving with Lorentz factor $\Gamma$ through a charge neutral
background, and if one defines the background plasma to be the shell,
then equation (14) gives a frequency of
\begin{equation}
   \omega^{\prime \, 2}
   = - { \beta^2 \left( \omega_{p, e,ism}^{\prime \, 2}
   + \omega_{p, i, ism}^{\prime \, 2} \right)
   \over
   \Gamma \left[ 1 + c^{-2} k_{\perp}^{-2}
   \left( \omega_{p, e, shell}^{\prime \, 2}
   + \omega_{p, i, shell}^{\prime \, 2} \right) \right] }
   \, .
\end{equation}
In this equation, $\omega_{p, e,ism}^{\prime}$
and $\omega_{p, i, ism}^{\prime}$ are the electron and ion plasma
frequencies of the interstellar medium,
and $\omega_{p, e, shell}^{\prime}$ and $\omega_{p, i, shell}^{\prime}$
are the electron and ion plasma frequencies of the shell, all measured
in the shell rest frame.
The negative value of $\omega^{\prime \, 2}$
shows that the wave grows.  Because
$\omega_{p, e, ism}^{\prime} > \omega_{p, i, ism}^{\prime}$
and $\omega_{p, e, shell}^{\prime} > \omega_{p, i, shell}^{\prime}$,
one finds that the growth rate of the filamentation of the electrons is
\begin{equation}
   \gamma_{fe}^{\prime} = { \beta \omega_{p, e, ism}^{\prime}
   \over \Gamma^{1 \over 2}
   \sqrt{ 1 + c^{-2} k_{\perp}^{-2} \omega_{p, e, shell}^{\prime \, 2} } }
   \approx {1 \over 2} \, \sqrt{ { 4 \pi e^2 \over m_e } }
   n_{ism}^{{1 \over 2}} \, .
\end{equation}
The growth rate is independent of wave length for
$k_{\perp} > \omega_{p,e,shell}^{\prime}/c$,
and it is $\propto k_{\perp}$ otherwise.  The length scale of the filament
is therefore given by $x_f = k_{\perp}^{-1} = c/\omega_{p, e, shell}$.
For ions alone, the growth rate, which differs from equation (16) only
in the numerator, is given by
\begin{equation}
   \gamma_{fi}^{\prime} = { \beta \omega_{p, i, ism}^{\prime} \over
   \Gamma^{1 \over 2}
   \sqrt{ 1 + c^{-2} k_{\perp}^{-2} \omega_{p, e, shell}^{\prime \, 2} } }
   \approx {1 \over 2} \, \sqrt{ { 4 \pi e^2 \over m_p } }
   n_{ism}^{{1 \over 2}} \, .
\end{equation}
This has the same length scale as the electron filamentation instability,
but a lower growth rate.

The filamentation instability growth rate is faster than the two stream
instability by the factor of $\Gamma$.  For $\Gamma = 10^3$,
$n_{ism} = 1 \, \cm^{-3}$,
one finds an electron filamentation growth rate of
$\gamma_{f, e}^{\prime} = 6 \times 10^4 \, \s^{-1}$,
which gives a length scale of $10^{6} \, \cm$ over which the instability
occurs.  For ions, the filamentation instability grows at a rate that is
a factor of $\sqrt{m_e/m_p}$ slower, so that for the parameters given above,
$\gamma_{f, p}^{\prime} = 1.4 \times 10^3 \, \s^{-1}$.  The growth rates
of all three instabilities are shown in Figures 3 and 4 as functions
of $\Gamma$.  Both filamentation
growth rates are therefore higher than the two-stream growth rate for
electrons using the parameters derived above for gamma-ray bursts.

\begin{figure}
\figurenum{3: Timescales}
\epsscale{0.7}
\plotone{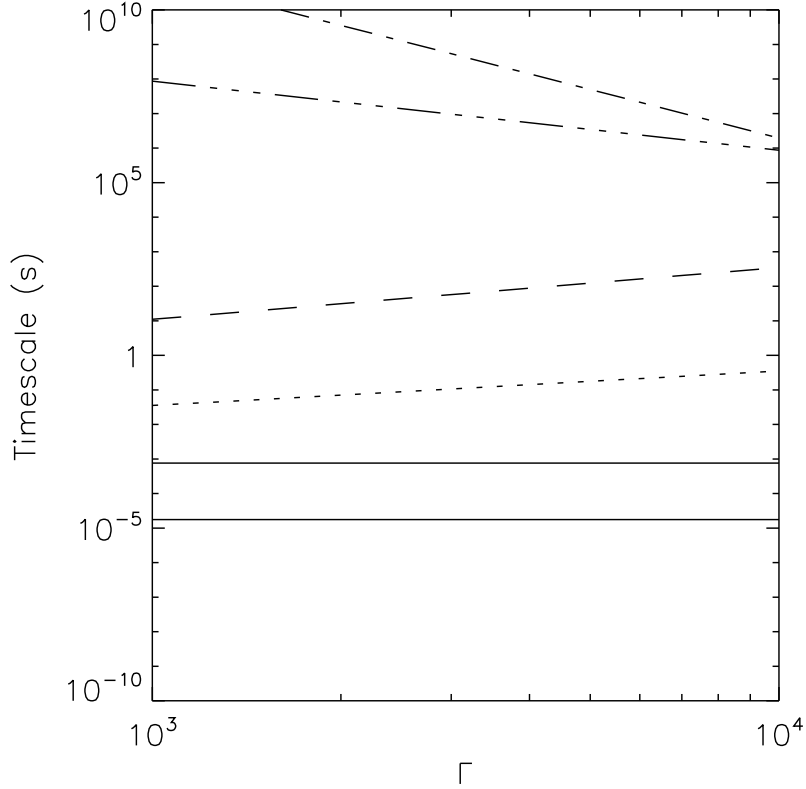}
\caption{Instability, radiative, and isotropization
timescales versus Lorentz factor.
Plasma timescales are given for an ISM plasma density of
$n_{ism} = 1 \cm^{-3}$. The lower two solid curves are the
linear growth timescales of the electron and the ion filamentation
instabilities, $t_{fe}^{\prime}$ and $t_{fi}^{\prime}$.
The dotted curve is the timescale for the electron two stream instability.
The short dashed curve is the timescale for the electron  distribution
to isotropize, $t_{iso}$.  The synchrotron cooling timescale 
$t_{sync}$ is given by the line with two dots and one dash, while the
synchrotron self-Compton cooling timescale $t_{Comp}$ is given by the
dash and dot line.}
\end{figure}

\begin{figure}
\figurenum{4: Timescales}
\epsscale{0.7}
\plotone{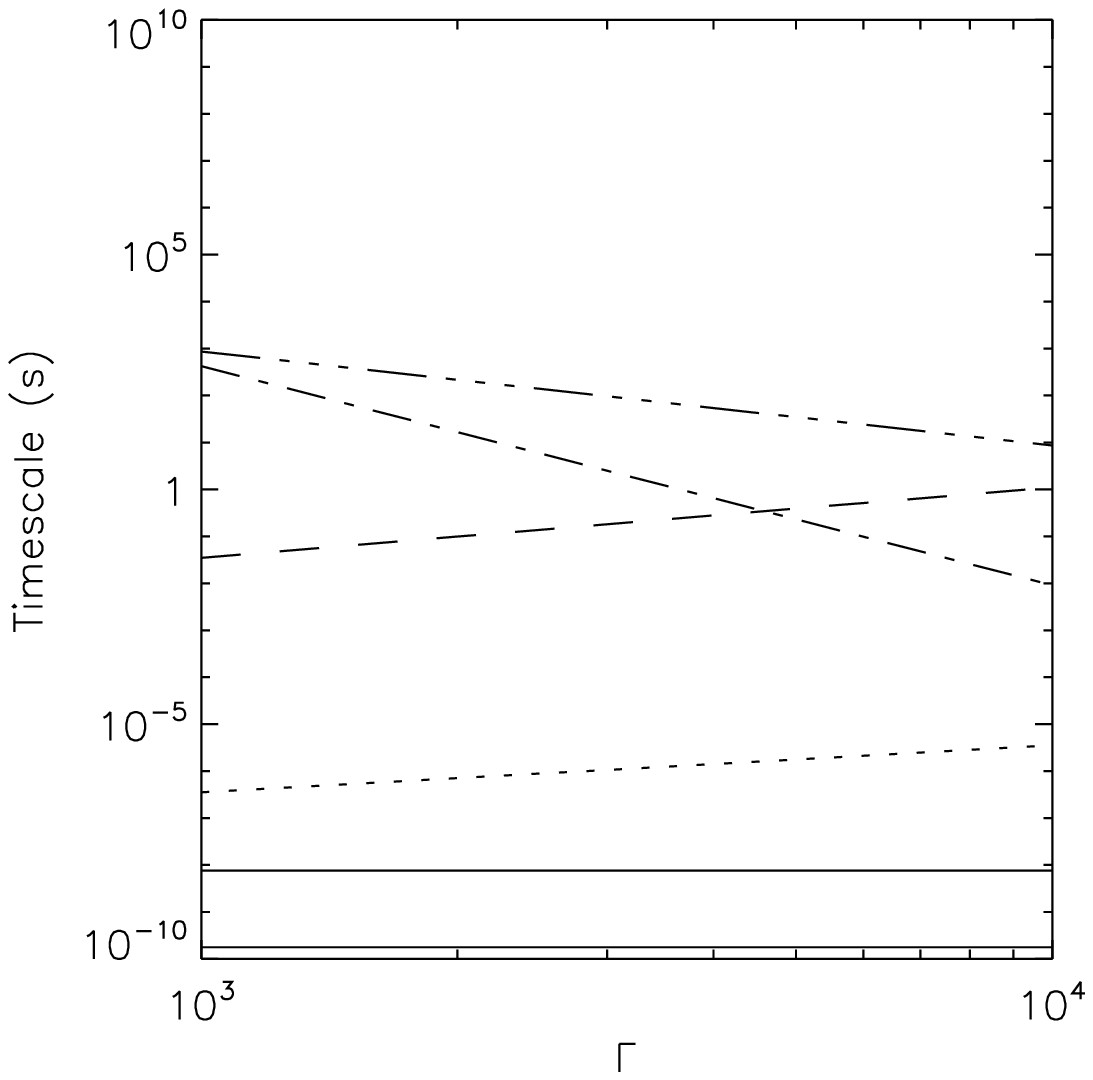}
\caption{Instability, radiative, and isotropization
timescales versus Lorentz factor. As in Fig\hbox{.}~3,
but for $n_{ism} = 10^{5} \cm^{-3}$.}
\end{figure}


\section{Filament Saturation}

The filaments grow until the growth rate of the thread equals the magnetic
bounce frequency of the particle beam producing the instability
(\markcite{Davidson2}Davidson \etal\ 1972;
\markcite{Lee}Lee \& Lampe  1973). 
The bounce frequency, which describes the motion of a particle across
the filament through the toroidal magnetic field,
is given by
\begin{equation}
   \omega_b = \sqrt{ e B / m c \Gamma x_f }
   \, ,
\end{equation}
where $x_f$ is the length scale of the filament,
and $m$ is the mass of the particles comprising the filament.
The size of the filaments is set by the lower limit on the wave number
for which growth occurs, $k_{\perp} c = \omega_{p,e, shell}^{\prime}$.
The ratio of the magnetic field energy density to the particle beam
energy density in the thread is then [35]
\begin{equation}
   { W_{B}^{\prime} \over W_{ism}^{\prime} }
   = { m_e n_{ism} \over m n_{shell}^{\prime} }
   = { m_e \over m  } \, \eta
   \, .
\end{equation}
One point to note about this equation is that
electron and proton components each generate magnetic fields of
the same strength.  For both components, one has
$W_{B}^{\prime} = m_e c^2 n_{ism} \eta \Gamma^2$, which,
for $\Gamma = \eta^{-1} = 10^3$,
gives $B^{\prime} = 0.14 \, G$ with $n_{ism} = 1 \, \cm^{-3}$,
and $B^{\prime} = 45.36 \, G$ with $n_{ism} = 10^5 \, \cm^{-3}$.
This independence is a consequence of the length scale of filamentation
being set by the shell electron density.

The thermalization of the ions within the thread can be estimated by
examining the equation of motion at saturation.  A particle's
equation of motion perpendicular to the thread for $u_x \ll \Gamma$ is
(\markcite{Davidson2}Davidson \etal\ 1972) 
\begin{equation}
   {d^2 x \over d \tau^2 } = - \omega_b^2 x
   \, .
\end{equation}
Solving this equation of motion using a maximum spatial amplitude of $x_f$,
one finds that the maximum momentum perpendicular
to the filament is $u_x = {x_f \omega_b \Gamma / c }$.
Replacing $x_f$ with the length scale of the filament, and
replacing the bounce frequency with the filament growth rate, one
finds
\begin{equation}
   u_x = \sqrt{ W_{B}^{\prime} \over W_{ism}^{\prime} } \Gamma \, .
\end{equation}
If $u_x \ll 1$, then the approximate energy that goes into thermalizing
the particles in the stream is approximately given by
$W_{th}^{\prime}/W_{ism}^{\prime} = u_{x}^2/\Gamma$,
while if $u_x \gg 1$, it is given
by $W_{th}^{\prime}/W_{ism}^{\prime} = u_{x}/\Gamma$.  As a result,
\begin{equation}
   { W_{th}^{\prime} \over W_{ism}^{\prime} }
   = \min\left( { m_e \over m } \, \eta \Gamma,
   \sqrt{ { m_e \over m } \, \eta } \, \right)
   \, .
\end{equation}
A point to note is that both of the terms on the right hand side
are greater than $W_{B}^{\prime}$, so that one always has
$W_{th}^{\prime} > W_{B}^{\prime}$.  Equations (19) and (22) are
plotted in Figure 5 for $m = m_p$ and $\Gamma = 10^3$ and~$10^4$.

\begin{figure}
\figurenum{5: Energy fractions}
\epsscale{0.7}
\plotone{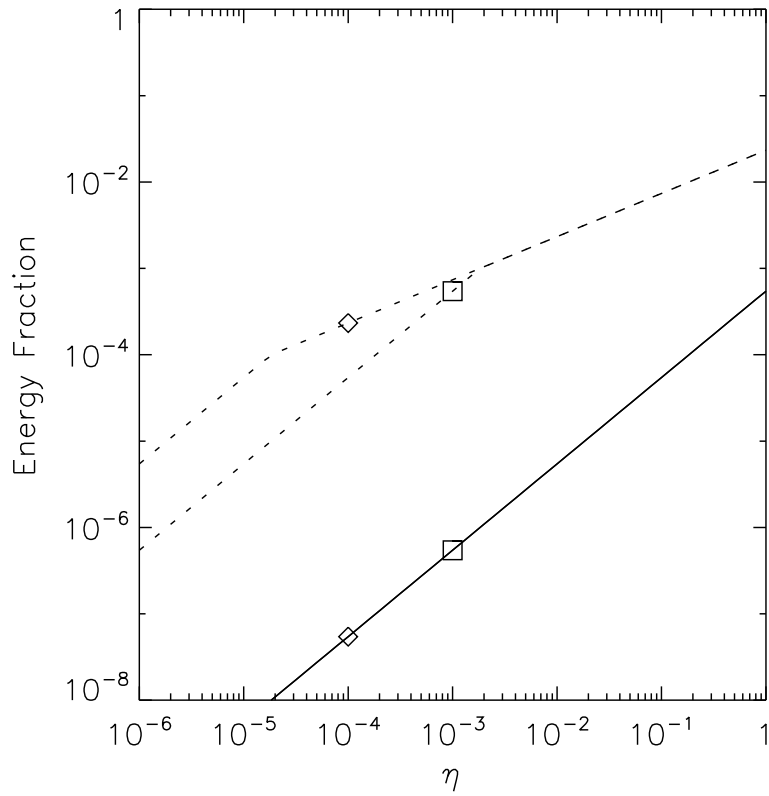}
\caption{The fraction of the ISM kinetic
energy, as measured in the shell rest frame, that goes into magnetic
field energy  (solid) and ion thermal energy (dotted) for $\Gamma = 10^3$
and $10^4$.}
\end{figure}

From these equations one sees that the filamentation instability
cannot mediate a shock.  If a shock were present, then $\eta$ would
be given by equation (2).
Placing this equation into equations (19) and (22) and
setting $m = m_p$, one finds that
${ W_{B}^{\prime} / W_{ism}^{\prime}} \approx { m_e / m_p \Gamma } \ll 1$
and
${ W_{i}^{\prime} / W_{ism}^{\prime}}
   = \left( { m_e / m_p \Gamma } \right)^{1/2} \ll 1$.
The energy released through the instability is small compared to the
kinetic energy of the interstellar medium as measured in the shell
rest frame.  This contradicts the Rankine-Hugoniot equations, and so
a shock never arises through the filamentation instability.


%
%

\section{Electron Thermalization}

Once ion filamentation is complete, electrons will attempt to come
into equilibrium through the two-stream instability.  This instability
will drive the electrons to a distribution that is uniformly distributed
in energy between the rest frame described by the ions in the filaments
and the ions in the shell.  The rest frame of the electron distribution
must preserve both the
electron charge density and the electron current, since the shell
plus interstellar medium is charge-neutral.  These two conditions
define a rest frame for the electrons that has a Lorentz factor
relative to the interstellar medium rest frame of
\begin{equation}
   \Gamma_e = { \Gamma + \eta \over \sqrt{ 1 + \eta^2 + 2 \eta \Gamma } }
   \, .
\end{equation}
Relative to this rest frame, the shell is moving with
\begin{equation}
   \Gamma_s^{\prime\prime} = { \eta \Gamma + 1
   \over \sqrt{ 1 + \eta^2 + 2 \eta \Gamma } }
   \, ,
\end{equation}
where the double primes are used to denote quantities measured in the
electron rest frame.

In the electron rest frame, the electron density is given by
$n_e^{\prime\prime} = n_{ism} \Gamma_e
+ n_{shell}^{\prime} \Gamma_s^{\prime\prime}$, which can be written as
\begin{equation}
   n_e^{\prime\prime} = n_{shell}^{\prime} \,
   \sqrt{ 1 + \eta^2 + 2 \eta \Gamma }
   \, .
\end{equation}
The electron rest frame has nearly the same Lorentz factor as the
shell rest frame as long as $\eta \ll \Gamma^{-1}$.  For
$1 \gg \eta \gg \Gamma^{-1}$, the Lorentz factor of the electrons
is $\Gamma_e = \sqrt{ \Gamma/2 \eta }$, which is a factor of
$1/\sqrt{ 2 \eta \Gamma}$ smaller than the Lorentz factor for the ions.
The value $\eta = \Gamma^{-1}$ is therefore an important transition
point for the character of the radiation emitted by the shell,
because the Lorentz boost of the radiation is smaller than the
boost associated with the shell when $\eta \gg \Gamma^{-1}$.
On the other hand, because $\eta < 1$,
$\Gamma_e^{\prime\prime} = \Gamma_e > \Gamma_s^{\prime\prime}$,
so the two-stream instability drives the electron distribution 
to the energy defined by $\Gamma_e$.

The electron in the shell rest frame has a Landau radius that is
much larger than the filament length scale.  Using the definitions for
$x_f$, the gyroradius, and equation (19) for the magnetic field
strength in the shell rest frame, one can write
\begin{equation}
   {r_e \over x_f} = u_e { m_e \over m_p } \eta
   \, .
\end{equation}
The gyroradius is therefore much larger than the filament width when
\begin{equation}
   u_e \gg {m_p \over m_e } \eta
   \, .
\end{equation}
For electrons thermalizing to the energy $\Gamma_e$,
$u_e \approx \Gamma/\sqrt{ 1 + 2 \eta \Gamma }$, so that equation (27)
becomes
\begin{equation}
   \Gamma \gg \cases{ {m_p \over m_e } \eta \, ,
   & if $\eta < \Gamma^{-1}$; \cr
   2 \left( {m_p \over m_e } \right)^2 \eta^3 \, ,
   & otherwise. \cr }
\end{equation}
One finds that the inequality holds for the upper term in equation (28)
whenever $\Gamma > \sqrt{m_p/m_e} = 42$; below we show that
$\Gamma \gtrsim 10^3$, so if equation (28) is to fail,
it will be for the lower term.  For $\eta \Gamma > 1$,
the inequality holds as long as
\begin{equation}
   \eta \Gamma < 2^{-{1 \over 3}} \,
   \left( { m_e \over m_p } \right)^{{2 \over 3}}
   \Gamma^{{4 \over 3}}
   = 52.9 \; \Gamma_3^{{4\over 3}} \, .
\end{equation}
When inequality (29) holds,
each electron passes through many filaments in a single orbit, and
the motion of the most energetic electrons
will be as a single fluid, but when the inequality fails, the electrons
will be confined to the local conditions within each filament, and the motion
of each electron will be determined by these local conditions.
For the remainder of the paper, we assume that equation (29) holds.

For the electrons to flow through the magnetic fields
generated in the shell, an average electric
field perpendicular to the magnetic field  must exist
in the shell rest frame. The electric
field is not uniform, since the system is charge neutral, but exists
only over distances of order the width of a filament, with values that are
proportional to the magnetic field strength.
When the gyroradius is large, the electron
effectively sees an average electric field as it completes one
orbit.  This electric field gives the electron guiding center a velocity
of $-U_s^{\prime\prime}$ in the shell rest frame.
The effective magnetic field strength
in the electron rest frame is then
\begin{equation}
   B^{\prime\prime} = {B^{\prime} \over \Gamma_s^{\prime\prime} }
   \, .
\end{equation}

Because the electron travels over many filaments,
the orientation of the magnetic field changes dramatically over one
gyroradius, so that the direction of the electron's velocity vector
is randomized through a random-walk process rather than through the
rotation over one orbit.  The electron travels the filament width
$x_f$ in the time $c \; \delta t^{\prime\prime}$.  In this time, the angle
$\theta \approx \delta t^{\prime\prime} /r_e$ is traveled,
where $r_e$ is the electron gyroradius.  Because the motion
is a random walk, the number of time intervals $\delta t^{\prime\prime}$
required to change direction by $2 \pi$ is approximately
$n = 4 \pi^2/\theta^2$.
The total amount of time required to isotropize the motion of the electrons
is therefore given by
\begin{equation}
   t_{iso}^{\prime\prime} = n \Delta t^{\prime\prime}
   = { 4 \pi^2 r_e^2 \over c \; x_f }
   = { 4 \pi^2 c \Gamma_e^2 \over \omega_c^{\prime\prime \, 2} x_f } \, ,
\end{equation}
where $\omega_c^{\prime\prime}$ is the cyclotron frequency.
Using equations (19) and (30) to define the magnetic field in equation
(31), one finds
\begin{equation}
   t_{iso}^{\prime\prime}
   = n \Delta t = { 2 \pi^2 \Gamma_{s}^{\prime\prime}
   \over \omega_{p,e,ism} \eta^{3/2} }
   = 11.1 \, \s \; n_{ism}^{-{1 \over 2}} \eta_{-3}^{-{3\over 2}}
   \, { \eta \Gamma + 1 \over \sqrt{ 1 + \eta^2 + 2 \eta \Gamma } }
   \, .
\end{equation}
For $\Gamma_{s}^{\prime\prime} \approx 1$, this
timescale is long compared to the two-stream instability for
electrons.


\section{Burst Radiation}

Two radiative processes are present within the theory: synchrotron emission
and Compton scattering.  The synchrotron emission will occur isotropically
in the electron rest frame describe by equation (23).  The observed
synchrotron radiation will be boosted into the observer's reference frame by
a factor of $\Gamma_e$.  Compton scattering of synchrotron radiation
by the synchrotron-emitting electrons
boosts the radiation by another factor of $\Gamma_e^2$,
because the electrons in this rest frame have a characteristic energy
of $m_e c^2 \Gamma_e$.

The characteristic synchrotron frequency in the shell rest
frame is given by
$h \nu^{\prime\prime} / m_e c^2 = 2 B^{\prime\prime}\Gamma_e^2/3B_{cr}$,
where $B_{cr} = e \hbar/m_e^2 c^3$.
Transforming this into the observer's reference frame and
using equations (19) and (30) to remove $B^{\prime\prime}$,
one finds that the characteristic synchrotron energy is
\begin{eqnarray}
   {h \nu_s \over m_e c^2}
   &= &{ 2 \sqrt{ 8 \pi} \hbar e \over 3 m_e^{3/2} c^2}
   \eta^{{1\over2}} n_{ism}^{{1\over 2}} \,
   { \Gamma_e^3 \Gamma \over \Gamma_s^{\prime\prime} }
   \, , \\
   &= &{ 2 \sqrt{ 8 \pi} \hbar e \over 3 m_e^{3/2} c^2 }
   \eta^{{1\over2}} n_{ism}^{{1\over 2}} \,
   { \left( \Gamma + \eta \right)^3 \Gamma
   \over \left( 1 + \eta^2 + 2 \eta \Gamma \right)
   \left( 1 + \eta \Gamma \right) }
   \, , \\
   &= &2.17 \times 10^{-6} \;
   \eta_{-3}^{{1\over 2}} n_{ism}^{{1\over 2}}
   { \Gamma_3^4 \over \left( 1 + 2 \eta \Gamma \right)
   \left( 1 + \eta \Gamma \right) }
   \, ,
\end{eqnarray}
where terms of order $\eta$ and higher were dropped in the last
equation.  Equation (35) places the characteristic energy
in the optical band for the given characteristic energies and
$\eta \Gamma < 1$.  As $\eta$ increases above $\Gamma^{-1}$,
the characteristic observed energy falls as $\eta^{-{3 \over 2}}$, so
$\eta = \Gamma^{-1}$ defines the maximum
characteristic frequency for
a given value of $\Gamma$.  To have a characteristic photon energy of
$m_e c^2$ at $\eta = \Gamma^{-1}$ requires
$\Gamma > 6.93 \times 10^4 \, n_{ism}^{-1/7}$.

The characteristic energy of the Compton emission after
a single scattering is a factor of
$\Gamma_e^2$ larger than the synchrotron characteristic energy, so
\begin{eqnarray}
   {h \nu_{C} \over m_e c^2} &= &{2 \sqrt{ 8 \pi} \hbar e \over 3 m_e^{3/2} c^2}
   \eta^{{1\over2}} n_{ism}^{{1\over 2}}
   { \left( \Gamma + \eta \right)^5 \Gamma
   \over \left( 1 + \eta^2 + 2 \eta \Gamma \right)^2
   \left( 1 + \eta \Gamma \right) } \, , \\
   &= &2.17 \; \eta_{-3}^{{1\over 2}} n_{ism}^{{1\over 2}}
   { \Gamma_3^6
   \over \left( 1 + 2 \eta \Gamma \right)^2
   \left( 1 + \eta \Gamma \right) } \, .
\end{eqnarray}
The characteristic energy of the Compton scattered radiation after a
single scattering is at
$m_e c^2$ when $\Gamma > 1.47 \times 10^3 \, n_{ism}^{1/11}$.  A second
scattering takes the photon in the observer rest frame
to the GeV energy range.
Further scattering does not change the photon energy, because the photon
energy is of order the characteristic
electron energy after the second scattering.

Each of the radiative components spans a broad range of energies.  The
low end of the synchrotron emission is set by the cyclotron frequency,
which is smaller than the characteristic synchrotron frequency by
a factor of $\Gamma_e^2$.  For $\eta \Gamma = 1$, the cyclotron frequency
is at $\nu \approx 2.68 \times 10^8 \, \Hz \; n_{ism}^{1/2} \Gamma_3^{3/2}$,
so that synchrotron emission extends down to the radio band.
An important point is that the cyclotron frequency is related to the
plasma frequency in the electron rest frame as
$\nu_{c}^{\prime\prime} \approx \sqrt{2 } \omega_{p,e}^{\prime\prime}
\eta \Gamma /2 \pi \left( 1 + 2 \eta \Gamma \right)^{1/4}$, so
that they are about equal, and the cyclotron photons escape the shell
to the observer.  The lowest energy of the Compton scattered
radiation is the cyclotron frequency photons upscattered by $\Gamma_e^2$,
which means that the low end of the Compton scattered radiation equals the
high end of the synchrotron emission.  If most of the energy is in the
Compton scattered component, and the synchrotron photon number
spectrum falls faster than $\nu^{-2}$, so that most of the energy is
released at the low end of the spectrum, then the low end of the
Compton spectrum will be larger than the high end of the synchrotron
spectrum.  This implies that optical and ultraviolet emission is part of
a single smooth continuum that extends through the x-ray and gamma-ray
bands, and that most of the energy emitted by the burst is released
in the optical and ultraviolet.

The ratio of the synchrotron emission rate to the single-scattering
Compton emission rate for a single electron is given by
$P_{sync}^{\prime\prime}/P_{c1}^{\prime\prime}
= W_{B}^{\prime\prime} / W_{synch}^{\prime\prime}$,
where $W_{B}^{\prime\prime}$ and $W_{synch}^{\prime\prime}$
are the magnetic field and the synchrotron
photon energy densities as measured in the electron rest frame.
The synchrotron energy density is related to the single electron
emission rate by
$W_{sync}^{\prime\prime}
   = P_{synch}^{\prime\prime} n_{emis}^{\prime\prime}
   l^{\prime\prime}/4 \pi c$,
where $n_{emis}^{\prime\prime}$ is the density of electrons with
Lorentz factor $\Gamma_e$ in the electron rest frame.
The synchrotron emission rate is given by
$P_{sync}^{\prime\prime} = 4 \sigma_T c \Gamma_e^2 W_{B}^{\prime\prime}/3$,
so the
ratio becomes
\begin{eqnarray} P_{sync}^{\prime\prime}/P_{c1}^{\prime\prime}
   &= &{ 3 \pi \left( \eta \Gamma + 1 \right)
   \over \sigma_T \left( \Gamma + \eta \right)^2 f_{emis} }
   \left( { 9 m_p \over {\cal M} n_{ism}^2 \Gamma^2 } \right)^{{1 \over 3}}
   \, , \\
   &= &3.50 \, \Gamma_3^{-{8\over 3}}
   { \left( \eta \Gamma + 1 \right) \Gamma^2 \over
   \left( \Gamma + \eta \right)^2 }
   n_{ism}^{-{2 \over 3}} {\cal M}_{27}^{-{1 \over 3}} f_{emis}^{-1}
   \left( { R \over R_0 } \right)^2 \, .
\end{eqnarray}

The ratio of the synchrotron emission to the rate at which energy
carried by the interstellar medium flows through the shell 
is given by
\begin{eqnarray} { {\dot E}_{sync}^{\prime\prime}
   \over m_p c^3 n_{ion} \Gamma_e^2 }
   &= &{4 m_e \sigma_T {\cal M}^{{1\over 3}}
   n_{ism}^{{2 \over 3}} \Gamma^{{8\over 3}} \eta
   \left( 1 + \eta \Gamma \right) f_{emis}
   \over 3^{{5 \over 3}} m_p^{{4 \over 3}} }
   \left( { R_0 \over R} \right)^2
   \, , \\
   &= &1.956 \times 10^{-6} {\cal M}_{27}^{{1\over 3}} 
   n_{ism}^{{2 \over 3}} \Gamma_3^{{5\over 3}}
   \left( \eta \Gamma \right) \left( 1 + \eta \Gamma \right)
   f_{emis}
   \left( { R_0 \over R} \right)^2
   \, .
\end{eqnarray}
For single scattering Compton cooling, this ratio is
\begin{eqnarray} { {\dot E}_{Comp}^{\prime\prime}
   \over m_p c^3 n_{ion} \Gamma_e^2 }
   &= &{2 m_e \sigma_T^2 {\cal M}^{{2\over 3}}
   n_{ism}^{{4 \over 3}} \Gamma^{{10\over 3}} \eta
   \left( \Gamma + \eta \right)^2
   f_{emis}^2
   \over 3^{{10 \over 3}} \pi m_p^{{5 \over 3}} }
   \left( { R_0 \over R} \right)^4
   \, , \\
   &= &2.80 \times 10^{-7} {\cal M}_{27}^{{2\over 3}} 
   n_{ism}^{{4 \over 3}} \Gamma_3^{{13\over 3}}
   \left( \Gamma \eta \right)
   { \left( \Gamma + \eta \right)^2 \over \Gamma^2 }
   f_{emis}^2
   \left( { R_0 \over R} \right)^4
   \, .
\end{eqnarray}
From these equations, one sees that the energy release is very inefficient
unless $n_{ism}$ or $\Gamma$ are larger than the characteristic values used
in the calculations.  We discuss this point in~\S 8.


\section{Radiative Timescales}


The cooling time scale for synchrotron emission of a relativistic
electron is found by dividing the electron energy $m_e c^2 \Gamma_e$ by
$P_{sync}^{\prime\prime}$,
the emissivity of a single electron.  In the electron rest frame, the
synchrotron cooling timescale is
\begin{eqnarray}
   t_{sync}^{\prime\prime} &= &{ 3 \over 4 \sigma_{T} c \eta n_{ism} }
   { \left( \eta \Gamma + 1 \right)^2 \over \Gamma^2
   \left( \Gamma + \eta \right) \sqrt{ 1 + \eta^2 + 2 \eta \Gamma } } 
   \, , \\
   &= &3.76 \times 10^{7} \, \s \;
   n_e^{-1} \, \Gamma_3^{-2} \,
   { \left( \eta \Gamma + 1 \right)^2 \over \eta
   \left( \Gamma + \eta \right) \sqrt{ 1 + \eta^2 + 2 \eta \Gamma } } 
   \, .
\end{eqnarray}
The synchrotron self-Compton cooling timescale is found
by multiplying equations (38) and (45) together, which gives 
\begin{eqnarray}
   t_{Comp}^{\prime\prime}
   &= &{ 9 \pi \over 4 \sigma_{T}^2 c f_{emitt} n_{ism} }
   \left( { 9 m_p \over n_{ism}^2 {\cal M} } \right)^{{1 \over 3}}
   { \left( \eta \Gamma + 1 \right)^3
   \over \eta \Gamma^{8\over 3}
   \left( \Gamma + \eta \right)^3 \sqrt{ 1 + \eta^2 + 2 \eta \Gamma } } \,
   \left( { R \over R_0 } \right)^2
   \, , \\
   &= &1.32 \times 10^8 \, \s \;
   {\cal M}_{27}^{-{1 \over 3}} \,
   n_{ism}^{-{5 \over 3}} f_{emitt}^{-1} \,
   \Gamma_3^{-{ 14 \over 3}} \,
   { \left( \eta \Gamma + 1 \right)^3 \Gamma^2 \over \eta
   \left( \Gamma + \eta \right)^3 \sqrt{ 1 + \eta^2 + 2 \eta \Gamma } }
   \left( { R \over R_0 } \right)^2
   \, .
\end{eqnarray}
These timescales are plotted in Figures 3 and 4 for
$n_{ism} = 1 \, \cm^{-3}$
and
$10^{5} \, \cm^{-3}$,
with
$\left( R/R_0 \right)^2 = m_p/m_e$, and $\eta \Gamma = 1$.
One sees that for the higher densities, the Compton cooling timescale
can fall below the timescale for isotropization.  When this occurs,
the radiative cooling will determine the shape of the electron distribution.
The Compton cooling timescale is shorter than the isotropization timescale
when
\begin{equation}
   n_{ism} > 1.16 \times 10^6 \cm^{-3}
   \left( { R \over R_0 } \right)^{{12 \over 7}}
   \Gamma_3^{-{37 \over 7}}
   \left[ { \eta^{{1 \over 2}} \Gamma^{{7 \over 2}}
   \left( \eta \Gamma + 1 \right)^2 \over \left( \Gamma + \eta \right)^3 }
   \right]^{{6 \over 7}}
   \, .
\end{equation}


\section{Observational Consequences}

The theory outlined above has within it two observational
selection effects that define lower limits on the values
of $n_{ism}$ and $\Gamma$.  These selection effects arise
because gamma-ray bursts are recognized as such through the
efficient emission of gamma-rays.

Because gamma-ray bursts are selected by their gamma-ray emission,
one must have a value of $\eta$ that is sufficiently small to give
gamma-rays through Compton scattering.  Because the largest value
of the photon energy occurs at $\eta = 1/\Gamma$, one can derive
a lower limits on $\Gamma$ by requiring the right hand side
of equation (37) be $> 1$ at this value of $\eta$:
\begin{equation}
   \Gamma_3 > 0.831 \; n_{ism}^{-{1 \over 11}}
   \, .
\end{equation}
The weak dependence on $n_{ism}$ implies that for all gamma-ray bursts,
the bulk Lorentz factor $\Gamma > 10^3$.  Events may occur with smaller
$\Gamma$, but these would emit in the optical and ultraviolet bands.

\begin{figure}
\figurenum{6: Characteristic radiation energies}
\epsscale{0.7}
\plotone{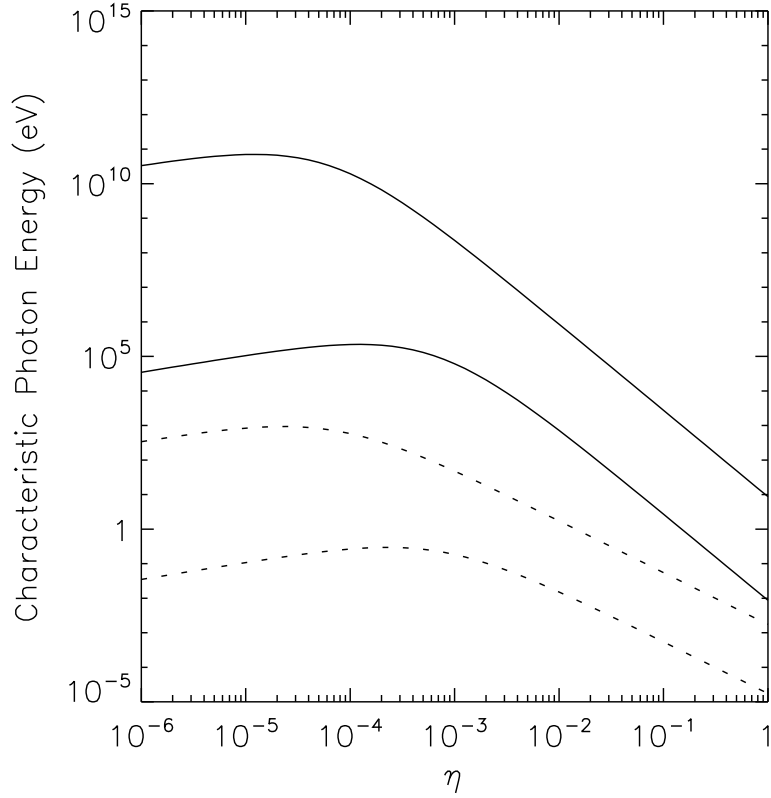}
\caption{The characteristic synchrotron emission (dotted lines)
and synchrotron self-Compton emission (solid lines)
energies of photons emitted from the relativistic
shell as measured by an observer at rest in the interstellar
medium for photons emitted along the velocity vector of the shell.   
Values of $\Gamma = 10^3$ (upper curves) and $10^4$ (lower curves)
are used.  The density is $n_{ism} = 1 \, \cm^{-3}$.}
\end{figure}

\begin{figure}
\figurenum{7: Characteristic radiation energies}
\epsscale{0.7}
\plotone{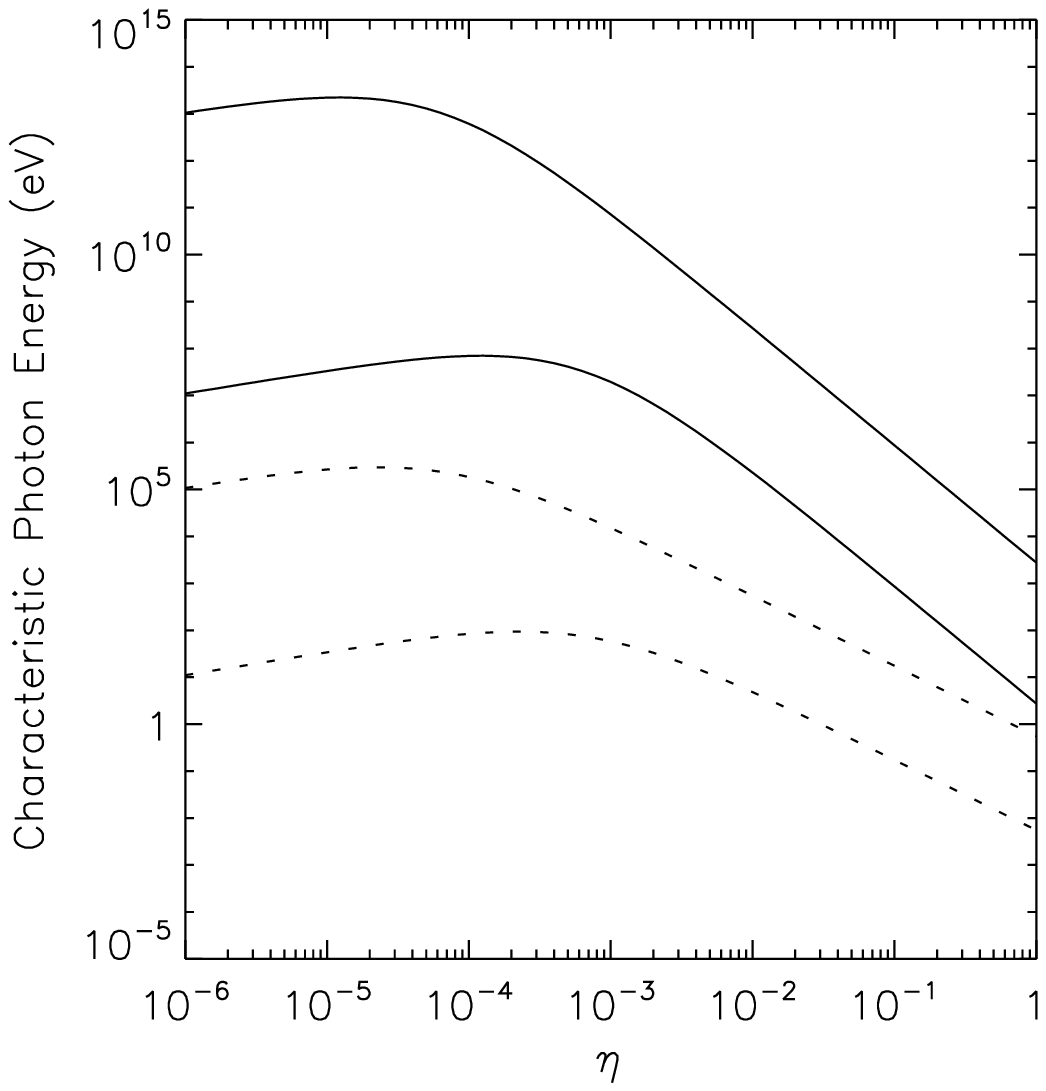}
\caption{Same as Fig\hbox{.}~6, but for $n_{ism} = 10^{5} \, \cm^{-3}$.}
\end{figure}

From Figures 6 and 7, one sees that an upper limit on the value of
$\eta$ is found from equation (37) for $\eta \Gamma \gg 1$.
This limit is
\begin{equation}
   \eta < \eta_{max} = 1.03 \times 10^{-3} \;
   n_{ism}^{{1 \over 5}} \, \Gamma_3^{{6 \over 5}}
   \, .
\end{equation}
These upper limits are plotted in Figures~1 and~2.  An important
aspect of these limits is that the timescales associated with the
shell thickness are generally $< 1 \, \s$.  For the higher density
figure, the time scales at $R/R_0 = 0.1$ range from
$0.01 \, \s$ to $1 \, \s$, which is consistent with
the shortest timescales exhibited by gamma-ray bursts.

Lower limits on the interstellar medium density are found by
requiring that Compton scattering efficiently remove energy
from the shell.  Two conditions must be met: first, Compton cooling
must dominate synchrotron cooling, and second, the Compton cooling
rate must be comparable to the rate at which energy is lost as
the shell decelerate over the distance $R$.  The first of these
conditions is derived from equation (39):
\begin{equation}
   n_{ism} > 
   6.54 \; {\cal M}_{27}^{-{1 \over 2}} \,
   f_{emis}^{-{3 \over 2}} \,
   { \left( \eta \Gamma + 1 \right)^{{3 \over 2}} \Gamma^3 \over
   \left( \Gamma + \eta \right)^3 } \, \Gamma_3^{-4} \,
   \left( { R \over R_0 } \right)^3 \, .
\end{equation}
The second of these conditions is found by equating the right
hand side of equation (43) to $\left(R_0/R\right)^{3} \, g$,
where $g \le 1$ is a measure of efficiency.  This gives
\begin{equation}
   n_{ism} = 8.22 \times 10^{4} \, \cm^{-3} \;
   {\cal M}_{27}^{-{1\over 2}} 
   \Gamma_3^{-{13\over 4}} f_{emis}^{-{3 \over 2}}
   \left({R \over R_0}\right)^{{3 \over 4}} \, g^{{3 \over 4}}
   \, .
\end{equation}
If $g$ is very small in equation (52),
then most of the energy lost by the shell to the interstellar
medium is not radiated away, making the burst dim and
unobservable.

\begin{figure}
\figurenum{8: Limits on density}
\epsscale{0.7}
\plotone{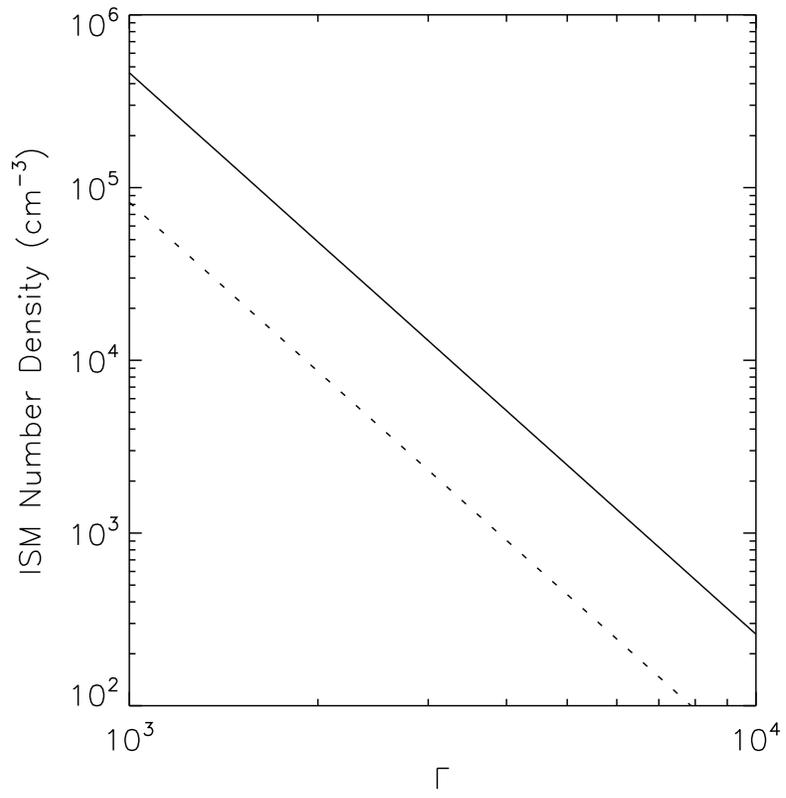}
\caption{The lower limits on $n_{ism}$
as a function of $\Gamma$.}
\end{figure}

The limits from equations (51) and (52) on $n_{ism}$
are plotted as functions of $\Gamma$ in Figure~8.
The point to note in this figure is
that the density required for high efficiency cooling is very high,
of order $10^{5} \, \cm^{-3}$.
This is important in explaining why all gamma-ray bursts are have a peak
value of the $\nu F_{\nu}$ curve near $250 \, \keV$.  The high source
density provides a medium for Compton attenuation of the burst spectrum.
Because the scattering medium is at rest in the galaxy rest frame, the value
of $E_p$ is independent of $\Gamma$, even though the characteristic
energy emitted by the shell is a strong function of $\Gamma$.  As a
consequence, if the medium density is low enough to keep attenuation from
occurring, then the density will be too low to efficiently produce gamma-ray
emission.  In such a circumstance, the shell will loose energy by
thermalizing ions and electrons, but the thermal energy of the
electrons will not be rapidly radiated away, making these particular
sources invisible.

An interesting consequence of the density on the theory is that as the
density increases, the timescale for radiative cooling becomes shorter than
the timescale for isotropization of the electron distribution.  The implications
of this is that there should be a coupling of the electron distribution
to the density of the interstellar medium, with the distribution falling
more rapidly, and being more anisotropic, at the higher
interstellar medium densities.  Because the shape of the
electron distribution determines the shape of the spectrum, one expects
the burst spectrum  to become softer for the higher interstellar medium
densities, and therefore for the higher attenuation optical depths.   This
is the case observationally, so this theory may provide an explanation
for that one characteristic of the Compton attenuation theory.

%
%
The theory as constructed has two natural timescales, one from the
thickness of the shell, and the second from
the deceleration distance $R$.  The shell thickness timescale is given
by equation (12), and is of order $2.9 \, \s$ for
$n_{ism} = 10^5 \cm^{-3}$ and $R/R_0 = 1$.  For the same
density and $R/R_0 = 10$, the timescale falls to $0.029 \, \s$.
The deceleration timescale is given by
\begin{equation}
   t_{R} = { R \over 2 c \Gamma^2}
   = 0.203 \, \s \; {\cal M}_{27}^{1/3} n_{ism}^{-1/3} \Gamma_3^{-7/3}
   \left( { R \over R_0 } \right)
   \, ,
\end{equation}
where the definition of $R_0$ in equation (5) has been used.
This timescale is less than the shell thickness
timescale whenever
\begin{equation}
   { R \over R_0 } > 8.74 \;
   \Gamma_3^{{2 \over 3}} \, \eta_{-3}^{{1 \over 3}}
   \, .
\end{equation}
Because ${ R \over R_0 }$ should be of order $m_p/m_e$,
the two timescales are of the same order.
When the two timescales are equal, the value is given by
\begin{equation}
   t_{R} = t_{shell} = 1.77 \, \s \;
   {\cal M}_{27}^{1/3} n_{ism}^{-1/3} \Gamma_3^{-5/3}
   \eta_{-3}^{{1 \over 3}} \, .
\end{equation}
For $n_{ism} = 10^5 \, \cm^{-3}$, the timescale is $0.038 \, \s$.
The timescales associated with both the shell and the deceleration
distance are sufficiently short to be responsible for the shortest
timescales observed in gamma-ray bursts.


\section{Summary of Conclusions}

To summarize the theory presented above, a baryonic shell with an
ultra-relativistic bulk velocity interacts with the interstellar medium
through the filamentation and the two-stream plasma instabilities.
The former instability gives rise to a magnetic field with a strength
that is far below the equipartition value, and the latter instability
heats the electrons to energies that are relativistic, but also far
below the equipartition value.  Neither instability is sufficient
to produce a shock.  Instead,
the interstellar medium passes through the shell, so that the region
behind the shell is not cleared of interstellar medium.
The electrons within the shell produce synchrotron radiation
with a characteristic energy in the observer's rest frame that ranges
from the radio to the ultraviolet.  Synchrotron self-Compton emission
by the electrons in the shell
produces x-rays and gamma-rays in the observer's rest frame, in addition
to optical emission.  The optical emission is dominated by the synchrotron
self-Compton component.  The timescales associated with the shell
thickness and the length scale over which the shell decelerates provide
a lower limit on the burst durations.  The burst duration itself would
be determined by the complex structure of the relativistic
wind, since the interstellar medium remains in place, permitting multiple
shells to each produce gamma-ray emission.

Two conditions must be met for the interaction between shell
and interstellar medium to efficiently produce gamma-rays:
first, the bulk Lorentz factor must be $> 10^3$ in order to
produce radiation above $1 \keV$; second, the number density
of the interstellar medium must be greater than
$\approx 10^6 \cm^{-3} \, \Gamma_3^{-{13 \over 4}}$, where
$\Gamma_3$ is the bulk Lorentz factor in units of $10^3$, for
the thermal energy to be radiated efficiently.  The lower limit
on $\Gamma$ through the selection
effect provides an explanation for why the value
of $\Gamma$ is always sufficiently high to allow the escape of
$1 \MeV$ photons from the gamma-ray burst emission region without
the production of an electron-positron plasma from photon-photon
pair creation and the subsequent thermalization of the radiation.
The limit on density provides an explanation of why all gamma-ray
burst spectra appear to be Compton attenuated.  Because the limits
on $\Gamma$ and $n_{ism}$ are from selection effects in observing
the emission of gamma-rays, one expects there to be burst events
with values of $\Gamma$ and $n_{ism}$ outside these limits.  For
bursts with low $\Gamma$, the bursts radiate at energies below
$1 \, \keV$.  Therefore, one expects a population of optical and
ultraviolet transients that have no gamma-ray emission.  For bursts
with low density, the radiation of energy is inefficient, so that
the bursts are of low intensity.  These bursts may appear in burst
samples through a correlation of burst intensity with the interstellar
medium density inferred from the Compton attenuation model.

An aspect of the theory that provides a test is the comparison of
instantaneous gamma-ray, x-ray, and optical emission to the radio
and optical emission.  There are two aspects of the theory
to test.  First, one can test whether the broad band spectrum
is consistent with being a synchrotron spectrum at low frequency
and a Compton scattered synchrotron spectrum at higher frequency.
Second, one can test the consistency of physical parameters in
the theory.  This last is done by comparing the Thomson and
photoelectric optical depths found through a fit of the Compton
attenuation model to the optical attenuation derived under the
assumption that the unattenuated gamma-ray continuum extends
to the optical band.  Third, if the optical spectrum is sufficient
to fix the cyclotron frequency by determining the low energy
drop-off of the Compton spectrum, one can solve for the value of
the Lorenz factor, which will provide a consistency test through
the lower limit on $\Gamma$.  If one can model the x-ray afterglow
of a burst as the forward scattering of x-rays by dust, then one has
additional information about the optical extinction that can be used
in this comparison.

The theory of afterglows in this theory has yet to be developed.
An important difference from the shock theory of afterglows is
that the region behind the shell will emit afterglow radiation
in competition with the radiation from the decelerated shell.
Because the evolution of the afterglow from the
interstellar medium is determined by the broadening of the
look-back surface and the radiative cooling of the interstellar
medium, while the evolution of the shell radiation is determined
by the decrease in $\Gamma$ and the evolution of the thermal
structure of the shell, the theory should have two distinct afterglow
components that produce a complex afterglow behavior.

Numerical modeling of the plasma processes can lead to additional
observational tests of the theory.  In particular, the
numerical modeling of the electron distribution for 
the regime where the electron isotropization timescale
exceeds the Compton cooling timescale (Fig.~2) may provide a
test through the correlation of Thomson optical depth with
spectral hardness.  One suspects that as $\Gamma$ increases,
the synchrotron spectrum becomes softer, because
an electron radiatively cools before it isotropizes, making the
electron distribution one-dimensional.  Because of the lower limit
on $n_{ism}$ in inversely related to $\Gamma$, one expects $n_{ism}$,
and therefore the Thomson optical depth, to be smaller for softer spectra.
This conjecture requires numerical verification; if it is verified,
then one can use the correlation of unattenuated spectral index with
Thomson optical as a test of the theory.  There is already some
evidence of such a correlation
(\markcite{Brainerd3}Brainerd \etal\ 1998, Fig. 8a).

Two theoretical investigations are now warranted.  The first is a study of
the broad band spectrum expected for this theory need to be numerically
calculated for a number of electron distributions.  This study will
determine what aspects of the spectrum provide tests of the theory
without requiring a full understanding of the plasma processes.  Such
a study can be carried out through Monte Carlo simulation, and should
include the effects of optical and Compton attenuation.  The goal is
to model the spectrum from the radio to the gamma-ray.  The second is
a study of the plasma processes.  This will require the development of
plasma codes to study the interactions of relativistic beams and the
growth of instabilities to the nonlinear regime.  Only such a study will
verify if the analytic conclusions reached above are accurate.
Only such a study will one determine the value of $\eta$
in terms of the other burst parameters.  And only
through such a study will one obtain model spectra that are dependent just
on the bulk Lorentz factor, the density of the interstellar medium, and the
mass of the relativistic shell.

The plasma instability theory is capable of explaining the most important
features of gamma-ray bursts.  Further theoretical
research is therefore justified, and
should lead to strong and unambiguous observational tests of the theory.


\end{document}